\documentclass[
    ,final            
  ]
  {aipproc}

						     \layoutstyle{6x9}

\begin{document}

\title{The initial conditions of the Universe and holography}


\author{Pablo D\'\i az}{
  address={Departamento de F\'\i sica Te\'orica\\
Universidad de Zaragoza. 50009-Zaragoza. Spain}
}

\author{Miguel Angel Per}{
  address={Departamento de F\'\i sica Te\'orica\\
Universidad de Zaragoza. 50009-Zaragoza. Spain}
}

\author{Antonio Segu\'\i }{
  address={Departamento de F\'\i sica Te\'orica\\
Universidad de Zaragoza. 50009-Zaragoza. Spain}
}

\begin{abstract}
We address the initial conditions for an expanding cosmology using the 
holographic principle. For the case of a closed model, the old prescription of 
Fishler and Susskind, that uses the particle horizon to encode the bulk 
degrees of freedom, can be implemented for accelerated models with enough 
acceleration. As a bonus we have singularity free bouncing models. The bound 
is saturated for co-dimension one branes dominated universes.
\end{abstract}

\maketitle


\section{Introduction}
Holography is one of the underlying principles of a full theory of quantum 
gravity {\cite{thooft}}.  This principle states that if the gravitational 
field is quantized, 
the degrees of freedom of a physical system can be mapped to other degrees of 
freedom of a dual theory now living on the boundary of the original one.
The paradigmatic realization of holography is the AdS/CFT correspondence 
\cite{Maldacena}. In the semiclassical regime, when the curvature scale 
is larger than the Planck length, the residual principle adopts the form 
of different sorts of entropy bounds. A characteristic of all of them is that the number of degrees of freedom of a physical system does not scale any more with the volume of the system but with the area that covers the system. This drastic reduction in the number of operative degrees of freedom is due to the formation of black holes. If we try to pack a given amount of information in smaller regions, because information is carried by energy, we pack an amount of energy in regions that in the limit can collapse to form a black hole; the information is lost at all effects.

The first time that the holographic principle was used to understand the FRW cosmological models was in \cite{susskind}; the prescription was to map the different degrees of freedom traversing the past light cone of the particle horizon (PH), on the PH area. Concretely, the prescription was to impose that the total amount of the entropy of the cosmic fluid traversing the past light cone of the PH was smaller than one quarter of the area of the PH at any cosmic time. For flat and open FRW universes, this covariant bound was effective; however, for closed spatial models, the bound is violated. The reason is that for closed models the PH, that at the beginning of the expansion grows, because the spatial sections are three spheres, at a given time begins to shrink diminishing its area, being less and less effective to store the increasing amount of entropy crossing its past light cone. In the limit, the area goes to zero as it reaches the antipodal point that constitutes a focusing point for the light cone. To solve this and other problems, Bousso introduced its covariant entropy bound prescription using the light cone as the recipient of the entropy and comparing this amount with the area of some \emph{preferred screens} that were appropriately defined \cite{bousso}. A crucial characteristic of the preferred screens was to define what constitutes its \emph{interior}. This characteristic was able to solve the problems previously commented for the Fischler Susskind entropy bounds on closed FRW models. 

Although the Bousso prescription is operationally effective for dealing with closed FRW models, its physical interpretation is not clear. When the PH traverses the apparent horizon, its interior changes side from the north hemisphere (where the big bang takes places) to the south one; this is because the interior is defined as the region where the null geodesic congruence orthogonal to the boundary has negative expansion. So, using Bousso's prescription we are relating the entropy that traverses the \emph{future} light cone to the area of the PH using an inversion of the arrow of time that seems not clear. We study in this communication the possibility of continuing the use of the PH as the depositary of the degrees of freedom carried by the cosmic fluid that traverses its \emph{past} light cone. This will restrict the nature of the cosmic fluid.

\section{Closed quintessence models are singularity free.}
 
In order to map the information traversing the past light cone of the PH onto itself, we need that the size of the PH never shrinks. To accomplish this, the expansion must be fast enough to prevent the horizon recolapse; in this way  the area of the PH never decreases. So we need to deal with accelerated expansions. The expansion is accelerated if $\ddot{R}>0$, where $R(t)$ is the scale factor of the universe. This occurs if $-1 \leq \omega<-1+2/n$, where $n$ is the number of spatial dimensions and $\omega$ appears in the equation of state of the cosmic fluid relating pressure and energy density $p=\omega \rho$.

It is interesting to note that the accelerated closed models do not present initial nor final singularities; in fact the relation between the scale factor $R$ and  the conformal cosmic time $d\eta=dt/R(t)$ as derived from the Friedman equations is
\begin{equation} 
R(\eta)=R_m \big(\sin \frac{n(1+\omega)-2}{2} \eta \big)^{\frac{2}{n(1+\omega)-2}},\label{factor}
\end{equation}
and presents two different regimes: if the exponent $\frac{2}{n(1+\omega)-2}$ is positive, which corresponds to a decelerated expansion, the scale factor (\ref{factor}) expands from an initial singularity at the big bang, attains a maximum value $R_m$ and then shrinks symmetrically until a final singularity is reached at the big crunch. If the exponent in (\ref{factor}) is negative,i.e. with values that corresponds to accelerated expansion, there is not initial, no final singularity and the arrow of time extends along the complete real line, describing a bouncing model, where the scale factor attains its minimum (bounce) at $R(t=0)=R_m$ given by
\begin{equation}
R_m=R_0 \big (\frac{1}{1-\Omega^{-1}_0} \big)^{\frac{1}{n(1+\omega)-2}};
\label{2}
\end{equation}
$R_0$ and $\Omega_0$ is the size and density of the universe at a given reference time $t_0$. This bouncing models are also known as oscillating models of second kind \cite{secondkind}.

We see that these accelerated models, needed to realize the Fischler Susskind entropy bound, has as a bonus that they are singularity free. However we will see that not all the accelerated models are able to give enough room in the PH to encode the amount of entropy traversing its past light cone; we need enough accelerated models that we study in the next section.

\section{The bound is saturated for co-dimension $1$ brane fluids}

We need to compare the area of the PH (measured in four times the Planck area) and the total flux of entropy that traverses the past light cone of the horizon. We suppose that the expansion is adiabatic and the evolution of the density of entropy is given by
\begin{equation}
s(t)=s_0 \frac{R_0^n}{R(t)^n} \label{3}.
\end{equation}
We also use the comoving coordinate $\chi$ given by
\begin{equation}
\chi= \int_0^r \frac{dr'}{\sqrt{1-r'^2}} \label{4},
\end{equation}
so that $r(\chi)=\sin (\chi)$. We rewrite (\ref{factor}) using $\alpha(n,\omega)$ defined by
\begin{equation}
\alpha(n,\omega)=\frac{n(1+\omega)}{n(1+\omega)-2},
\label{5}
\end{equation}
so that, because $\alpha<0$ (acceleration),
\begin{equation}
R(\eta)=\frac{R_{m}}{\big( \sin \frac{\eta}{1+| \alpha |} \, \big) ^{1+| \alpha |}} \, ,\label{6}
\end{equation}
where now 
\begin{equation}
 R_{m}=R_{0}\big( 1-\Omega_{0} \big) ^{\frac{2}{1+| \alpha |}} \, .\label{7}
\end{equation}
The \emph{ big bang} takes place for the smaller value for scale factor $R_m$, which can be arbitrarily small and that corresponds to a value of the conformal time given by $\eta _{BB}=\pi (1+|\alpha |) /2 $. The comoving coordinate of the PH will be
\begin{equation}
 \chi _{PH}(\eta )=\eta -\eta_{BB}=\eta -\frac{\pi}{2} (1+| \alpha |) \, .\label{8}
\end{equation}

The scale factor (\ref{6}) diverges for a value of the conformal time $\eta_{\infty}=(1+| \alpha |) \pi$. Using (\ref{8}), the value of the angle $\chi$ 
that asymptotically localizes the horizon is 
\begin{equation}
\chi _{PH}(\eta_{\infty})=\eta _{\infty}-\eta_{BB}=\frac{\pi}{2} (1+| \alpha |).\label{9}
\end{equation}
If we want to avoid the re-convergence of the light cone in the antipodal point, the value of $\chi_{PH}(\eta_{\infty})$ must be limited by $\pi$. As a consequence, the value of $\alpha$ must be greater than $-1$; note that this result is independent on the number of spatial dimensions $n$. Using (\ref{5}) we obtain
\begin{equation}
\omega < \frac {1}{n} -1.\label{10}
\end{equation}

The limiting case of $\alpha=-1$ has been also studied \cite{nosotros} obtaining, for the quotient between the entropy that traverses the past light cone of the horizon $S_{PH}$ and the area of this horizon $A_{PH}$, 
\begin{equation}
\frac{S_{PH}}{A_{PH}}(\eta)=\frac{s_{m}}{4} \ \tan ^{2} \frac{\eta}{2} \
( \eta- \pi - \sin \eta \cos \eta ),\label{11}
\end{equation}
where $s_m$, of order one in Planck units, is the reference spatial density of entropy in the bounce; the future infinity corresponds to $\eta_{\infty}=2 \pi$. The value (\ref{11}) remains finite on this range. Now we allow the PH to reach the antipodal point $\chi_{PH}(\eta_{\infty})=\pi$, but the divergence of the scale factor is stronger, giving rise a still infinite area of the PH.

We have also \cite{nosotros} studied the behavior of $\frac{S_{PH}}{A_{PH}}(\eta)$ near the bouncing time. Taking the limit of the general expression
\begin{equation}
\frac{S}{A}(\chi_{PH})=s_{m} \frac{\chi_{PH}-\sin \chi_{PH} \cos \chi_{PH}} {\sin ^{2} \chi_{PH}} (\cos \frac{\chi_{PH}}{1-\alpha})^{2(1-\alpha)} \label{12}
\end{equation}
for $\chi$ near the \emph{big bang}, (\ref{12}) remains finite approaching $2 s_m \chi_{PH}/3$.

It is interesting to note that the value that saturates the bound corresponds to the limiting value $\alpha=-1$ which, in terms of the parameter of the equation of state is $\omega=1/n-1$; such value corresponds to the relation between energy density ($\rho$) and pressure (minus tension) of a gas of co-dimension one branes \cite{lisa}.

\begin{theacknowledgments}
This work has been partially supported by MCYT (Spain) under grant FPA2003-02948.
\end{theacknowledgments}

\end{document}